\documentclass[twocolumn]{emulateapj}
\usepackage{times, graphicx}

\shorttitle{Unravelling the components of a multi-thermal coronal loop}
\shortauthors{Krishna Prasad et al.}
\newcommand\natphys{Nat. Phys.}

\begin{document}
\title{Unravelling the components of a multi-thermal coronal loop using magnetohydrodynamic seismology}
\author{S. Krishna Prasad} 
\affil{Astrophysics Research Centre, School of Mathematics and Physics, Queen's University Belfast, Belfast, BT7 1NN, UK.}                                  
\email{krishna.prasad@qub.ac.uk}
\author{D. B. Jess} 
\affil{Astrophysics Research Centre, School of Mathematics and Physics, Queen's University Belfast, Belfast, BT7 1NN, UK.}                                  
\affil{Department of Physics and Astronomy, California State University Northridge, Northridge, CA 91330, U.S.A.}                                  
\author{J. A. Klimchuk}
\affil{Heliophysics Division, NASA Goddard Space Flight Center, Greenbelt, MD, 20771, USA.}                                  
\author{D. Banerjee}
\affil{Indian Institute of Astrophysics, II Block Koramangala, Bengaluru 560034, India}

\begin{abstract}
Coronal loops, constituting the basic building blocks of the active Sun, serve as primary targets to help understand the mechanisms responsible for maintaining multi-million Kelvin temperatures in the solar and stellar coronae. Despite significant advances in observations and theory, our knowledge on the fundamental properties of these structures is limited. Here, we present unprecedented observations of accelerating slow magnetoacoustic waves along a coronal loop that show differential propagation speeds in two distinct temperature channels, revealing the multi-stranded and multi-thermal nature of the loop. Utilizing the observed speeds and employing nonlinear force-free magnetic field extrapolations, we derive the actual temperature variation along the loop in both channels, and thus are able to resolve two individual components of the multi-thermal loop for the first time. The obtained positive temperature gradients indicate uniform heating along the loop, rather than isolated footpoint heating.
\end{abstract}


\section{Introduction}
Bright arc-like structures extending from the Sun, when seen in Extreme-UltraViolet (EUV) wavelengths, are called coronal loops. Being a characteristic feature of the corona, it is important to assimilate high-resolution information on these structures in order to understand how plasma in the outer atmosphere is heated to multi-million Kelvin temperatures. The plasma temperatures and pressures are known to vary along the lengths of the loops, whose gradients provide some insight on the underlying heating mechanisms \citep{1978ApJ...220..643R}, but the main attribute of critical importance is the cross-field morphology \citep{2007ApJ...667..591P}. It has been suggested that coronal loops consist of multiple unresolved thin strands \citep{2014LRSP...11....4R,2015RSPTA.37340256K}. Observational evidence demonstrating low filling factors \citep{1999A&A...342..563D}, different degrees of `fuzziness' observed in loops at different temperatures \citep{2009ApJ...694.1256T,2011ApJ...736L..16R}, and plasma dynamics at small spatial scales \citep{2012ApJ...745..152A} indicate that this scenario is likely to be true. The typical cross-sections of the strands were estimated to be a few tens to a few hundreds of kilometres \citep{2015RSPTA.37340256K,2012ApJ...745..152A,2007ApJ...661..532D,2013ApJ...772L..19B}. Possible braiding of these strands can instigate magnetic reconnection, thus releasing significant energy to directly heat the plasma inside the loop \citep{2015RSPTA.37340256K,2013Natur.493..501C}. Each strand is believed to be impulsively and independently heated, leading to a multi-thermal configuration across the loop \citep{1994ApJ...422..381C,2005ApJ...633..489R,2008ApJ...682.1351K,2011ApJ...740..111T}. However, not all studies are consistent with this scenario since some loops demonstrate isothermal structuring \citep{2003A&A...406.1089D,2008ApJ...674.1191N,2009ApJ...691..503S}. One of the main issues plaguing observations of coronal loops is the intrinsic optically thin emission \citep{2003A&A...406.1089D,2010A&A...515A...7T}. Since the observed intensities are integrated along the line-of-sight, it becomes difficult to exclude emission from overlapping structures, which can be formed at different temperatures and inadvertently imply a multi-thermal structure. In this letter, we use a unique set of observations to reveal the multi-stranded and multi-thermal nature of a coronal loop, and isolate two of its components through the application of MHD seismology. More importantly, our results are free from the adverse effects arising from line-of-sight integrations. 

\section{Observations}
The Atmospheric Imaging Assembly \citep[AIA;][]{2012SoPh..275...17L,2012SoPh..275...41B} onboard the Solar Dynamics Observatory \citep[SDO;][]{2012SoPh..275....3P} captures full-disk images of the Sun at high spatial resolution ($\approx$870 km) nearly simultaneously in 10 different wavelength channels in the visible-EUV range of the electromagnetic spectrum. Images of a nearly circular sunspot, part of active region NOAA 11366 acquired by AIA on 10 December 2011 through two of its EUV channels centred at 171{\,}\AA\ and 131{\,}\AA, constitute the data used in this study. The time sequence starting from 15:30 UT until 18:00 UT is considered which comprises 750 images per channel at a cadence of 12~s. All of the data were prepared following standard procedures. Images in each channel were co-aligned to the first image using intensity cross-correlation. The peak temperature sensitivities of the 171{\,}\AA\ and 131{\,}\AA\ channels are at 0.7 MK and 0.4 MK \citep{2012SoPh..275...17L,2010A&A...521A..21O}, respectively. Note that the 131{\,}\AA\ channel is also able to observe high-temperature ($\approx$10 MK) components, such as those found in flaring plasma \citep{2012SoPh..275...17L}. However, the high-temperature components can be safely neglected for the present region due to the non-flaring environment. 

\section{Analysis and Results}
The location of the selected sunspot and a subfield enclosing it are shown in Fig.~\ref{fig1}a. The 171{\,}\AA\ and 131{\,}\AA\ images of this region show coronal loops extending from within the spot umbra to about 20 -- 30 Mm away in almost all directions, forming an extended fan-like structure (see Figs.~\ref{fig1}b-\ref{fig1}e). The images, when observed as time-lapse movies, display propagating features along these loops. Fourier analysis on lightcurves along the propagation path reveal narrow peaks around 6~mHz illustrating the prevalence of $\approx$3~min oscillations in the data. To enhance the visibility of propagating features, we employed Fourier filtration of the time series at each spatial position to retain only oscillations manifesting within a narrow window between 2 and 4 min. The filtered image sequences along with some sample (unfiltered) Fourier power spectra along the propagation path are available online. These sequences clearly show alternate bright and dark fronts moving outwards, which are indicative of propagating compressive waves. A debate is in progress on the possible ambiguity caused by the high-speed quasi-periodic upflows \citep{2012RSPTA.370.3193D,2015SoPh..290..399D}, which show similar signatures. However, the sunspot oscillations observed here, with wave fronts propagating symmetrically in all directions, unequivocally represent slow magneto-acoustic waves. It may be noted that these waves are also observed in other coronal channels (for e.g., AIA 193{\,}\AA). However, smaller amplitudes, contributions from lower temperature lines to the 193{\,}\AA\ passband \citep{2011A&A...535A..46D}, which could be as high as 30 -- 40\% of the total emission \citep{2012SoPh..279..427K}, make them unsuitable for the present study.

\subsection{Accelerating slow magnetoacoustic waves}
\begin{figure}
\centering
\hspace*{0.25in}
\includegraphics[scale=0.7, clip=true]{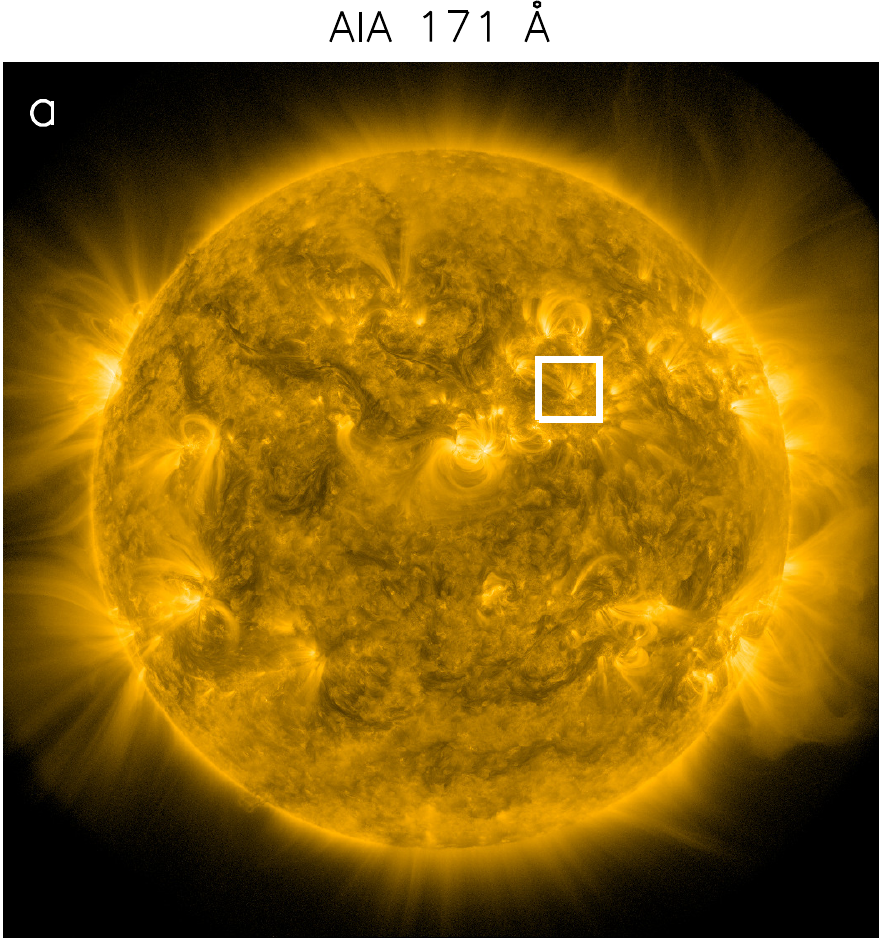}
\hspace*{0.1in}
\includegraphics[scale=0.55, clip=true]{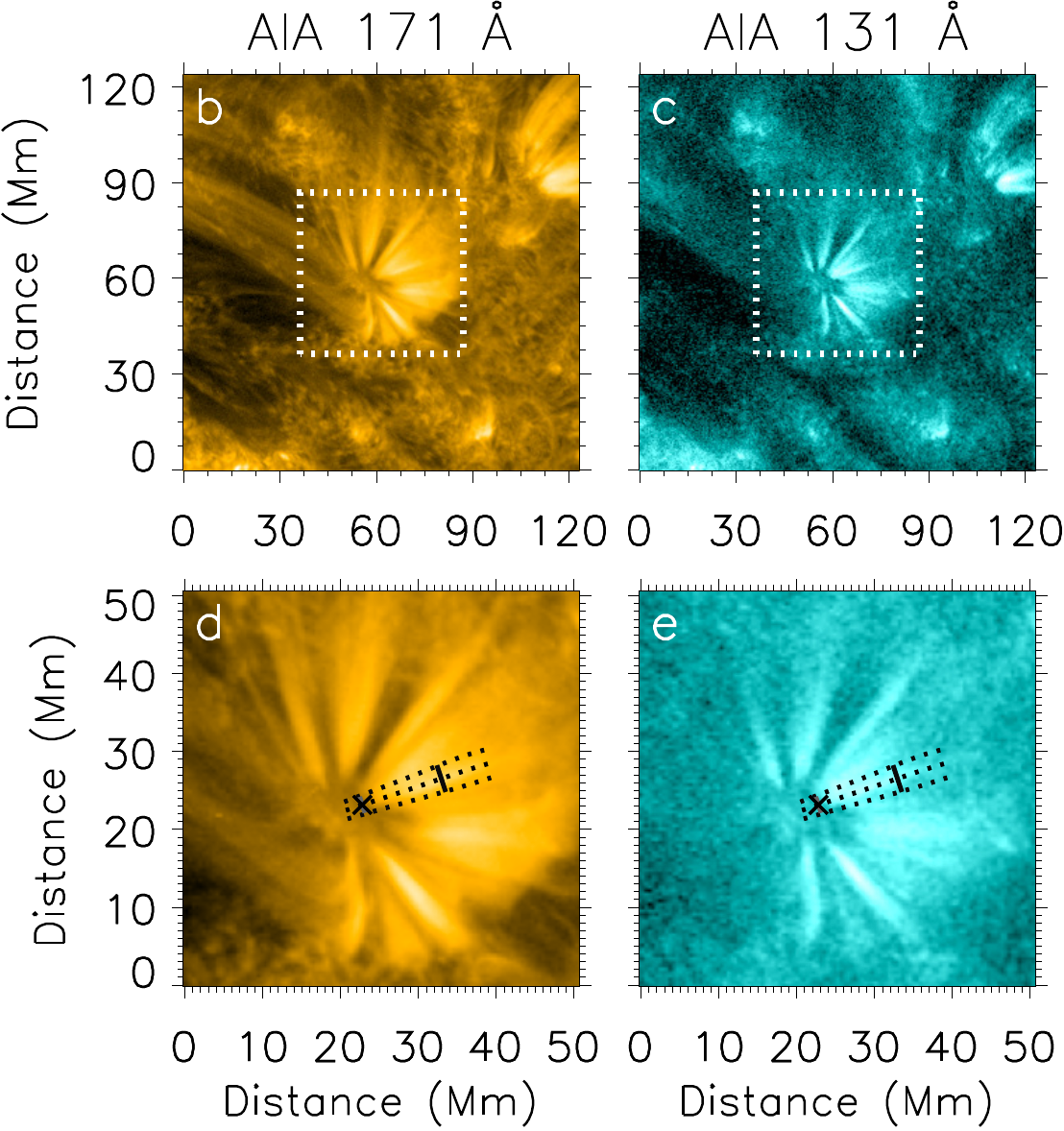}\\
\vspace*{0.1in}
\includegraphics[scale=0.5, clip=true]{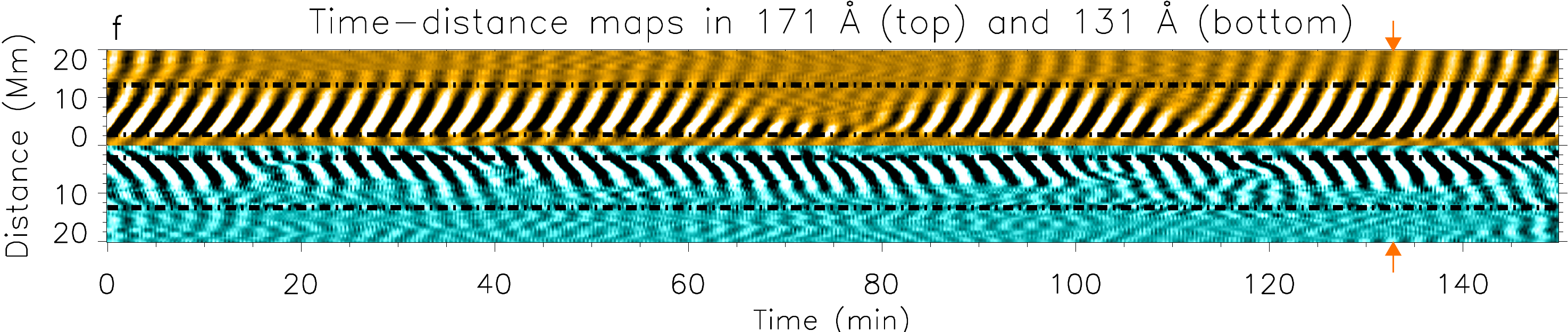}
\caption{a) A snapshot of the full-disk Sun captured by the AIA 171{\,}\AA\ EUV channel on 10 December 2011 at 15:30 UT. The white box outlines a 120$\times$120 Mm$^{2}$ region surrounding the sunspot under investigation. b), c) Subfields showing the vicinity of the sunspot in 171{\,}\AA\ and 131{\,}\AA\ channels. White dashed boxes outline the region used in the present analysis. d), e) Close up view of the sunspot. Dotted lines show the location of the track chosen along the loop for time-distance analyses. The central line follows the spine of the loop, while the lines on either side mark the region averaged during the time-distance analyses. Solid lines drawn across the loop bounds the section where propagating waves displayed adequate signal. The black cross identifies the location of the loop footpoint where field extrapolations are examined. f) Time-distance maps (after Fourier filtration) in the 171{\,}\AA\ and 131{\,}\AA\ channels constructed from the tracks shown in d) and e). The horizontal dash-dotted lines correspond to the locations of solid lines drawn across the loop in panels d) and e). Further analysis is restricted to the region enclosed by these lines. The red arrows mark the temporal location where spatial damping has been presented in Fig.~\ref{fig4}. A movie displaying evolution of the region shown in panels d) and e) is available online.}
\label{fig1}
\end{figure}

\begin{figure}
\centering
\includegraphics[scale=0.8, clip=true]{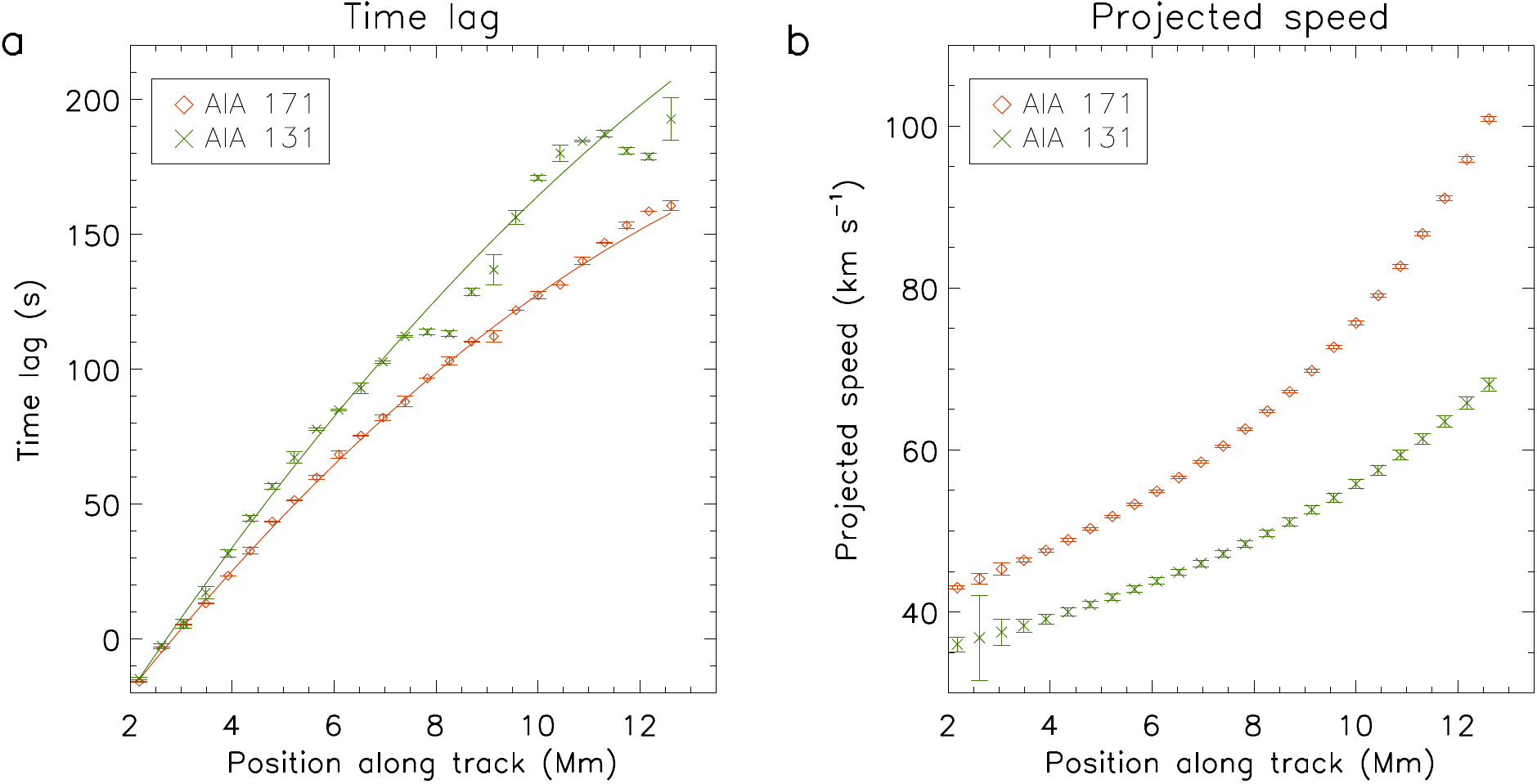}
\caption{a) Time delays measured at different positions along the track (see Fig.~\ref{fig1}) following the cross-correlation technique. A three-row average at the bottom is taken as the reference. Solid lines represent a second order polynomial fit to the data. b) Projected propagation speeds as a function of distance along the track for the 171{\,}\AA\ and 131{\,}\AA\ channels. The derivatives of the fitted values in a) are used to estimate these values, with the corresponding measurement errors also displayed.}
\label{fig2} 
\end{figure}

To study the observed waves in detail, we tracked their path along one of the loops (see Fig.~\ref{fig1}) and constructed time-distance maps \citep[\textit{e.g.}][]{bc1999SoPh186,2000A&A...355L..23D,2012SoPh..281...67K} for both temperature channels. Again, the time series at each spatial location is Fourier filtered to retain only oscillations between 2 and 4 min. The final maps, following Fourier filtration, are shown in Fig.~\ref{fig1}f, which display alternating bright and dark ridges representing outwardly propagating slow waves. In these maps, the footpoints along the track are joined together in the middle, with distances along each path increasing vertically outwards from the center, hence displaying a typical `fishbone' pattern \citep{2003A&A...404L...1K}. The ridges embedded within each channel commence simultaneously, indicating that the loop structures observed at these two temperatures are one and the same. The inclinations of the ridges provide a measure of the propagation speeds. As can be seen, the ridges in both channels display non-constant inclinations, suggesting acceleration and the subsequent increases in their propagation speeds. To further quantify this, we measure time lags at each spatial position using a cross-correlation technique \citep{2009ApJ...697.1384T}. We restrict this estimation to the region bounded by the dash-dotted lines displayed in Fig.~\ref{fig1}f. The lines in the center mark the locations of the loop footpoint (Figs.~\ref{fig1}d, \ref{fig1}e), while the outer lines mark the distances up to which the signal was deemed reliable. 
The measured time lags and the associated errors are plotted in Fig.~\ref{fig2}a. Solid lines represent a second-order polynomial fit to the data. An inverse derivative of the fitted values was then used to estimate the propagation speeds that are displayed in Fig.~\ref{fig2}b, alongside the respective errors. These values clearly show acceleration in both 171{\,}\AA\ and 131{\,}\AA\ channels. 

\subsection{Inferring the thermal structure}
\begin{figure}
\centering
 \includegraphics[scale=1, clip=true]{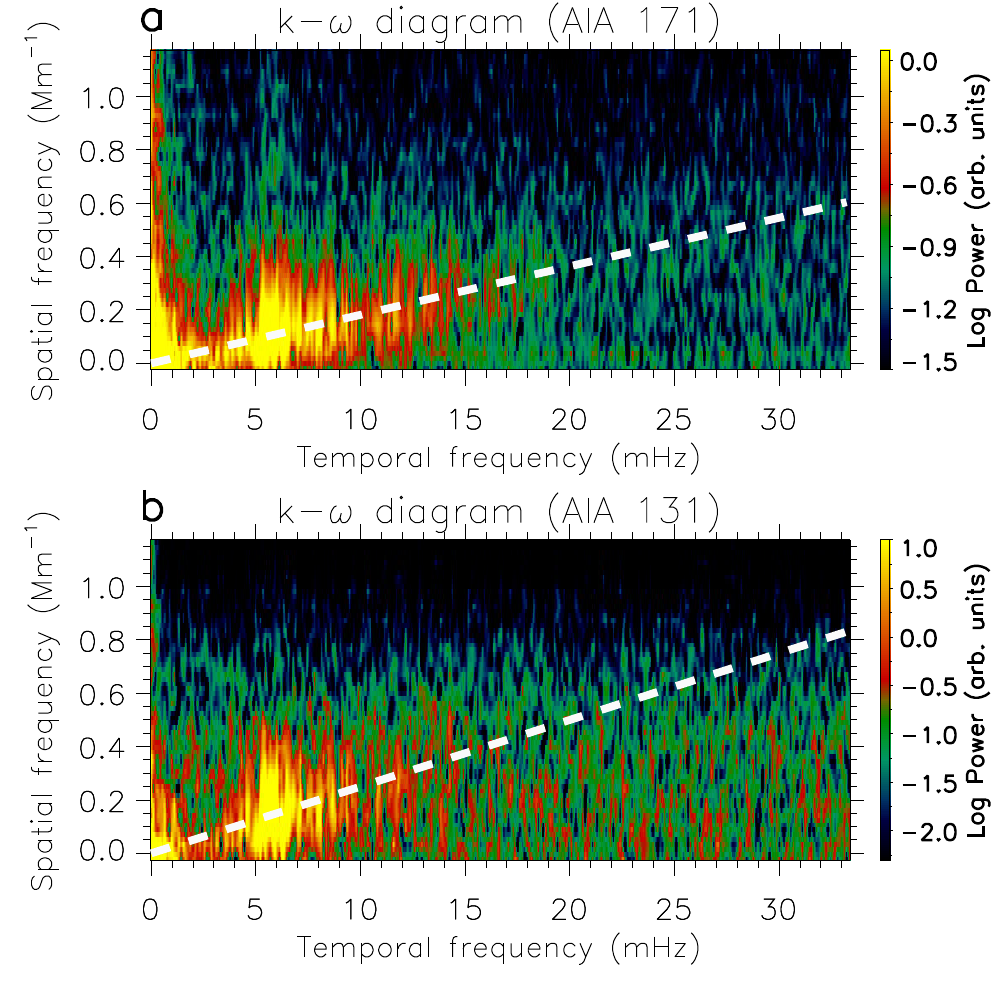}
 \caption{a) $k$-$\omega$ diagram generated from the original AIA 171{\,}\AA\ time-distance map, for the outward propagating waves. The white dashed line corresponds to a propagation speed of 55 km{\,}s$^{-1}$. b) Same as a) but for AIA 131{\,}\AA\ channel. The white dashed line corresponds to a propagation speed of 40 km{\,}s$^{-1}$. }
\label{fig3}
\end{figure}

Figure~\ref{fig3} displays the $k$-$\omega$ diagrams for the AIA 171{\,}\AA\ and 131{\,}\AA\ channels corresponding to the outward propagating waves, as described in \cite{2009ApJ...697.1384T}. These diagrams are generated from the initial (unfiltered) time-distance maps. Higher oscillatory power is clearly evident along a ridge, indicating a linear relation between the frequency, $\omega$, and the wavenumber, $k$. This behavior further confirms that the observed waves follow the dispersion relation for slow magneto-acoustic waves, $\omega=vk$, where $v$ is the propagation speed. The white dashed lines in Figs.~\ref{fig3}a \& \ref{fig3}b corresponds to a speed of 55 and 40~km{\,}s$^{-1}$, respectively. One may note that the 
ridges in the k-omega diagrams are rather wide in the vertical direction. This is a consequence of the spread in the observed propagation speeds due to acceleration. Hence the white dashed lines in Fig.~\ref{fig3} only provide an indication of the range of phase speeds present in the data.

\begin{figure}
\centering
\includegraphics[scale=0.55, clip=true]{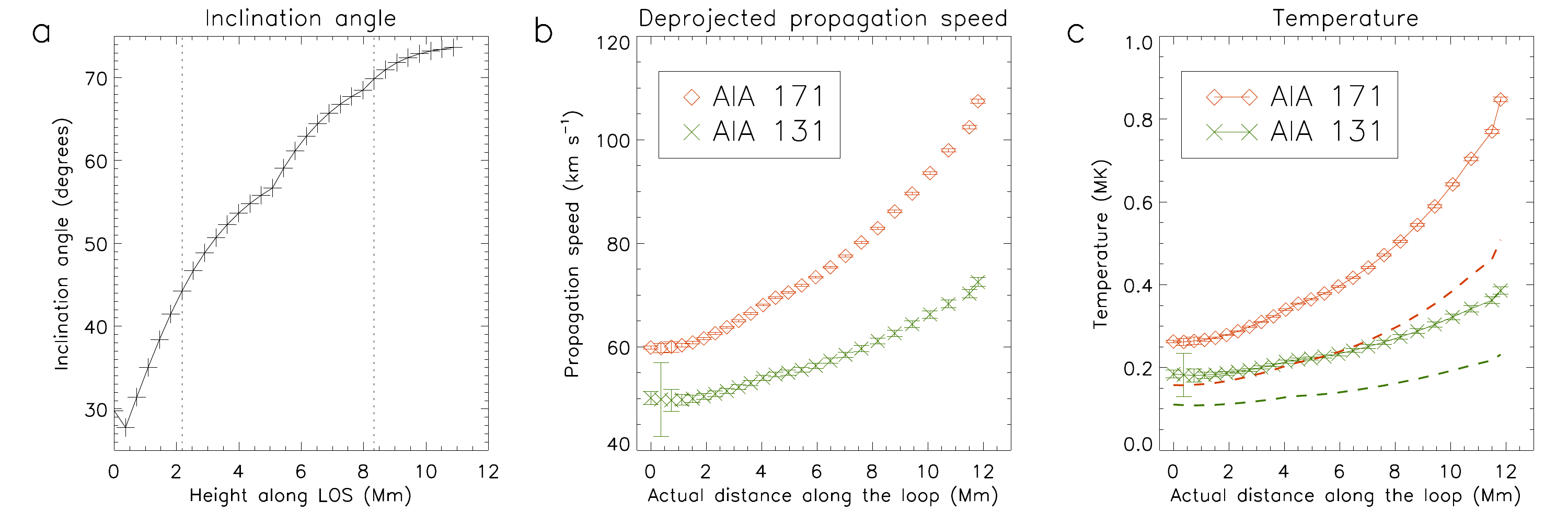}
\caption{a) Inclination of the loop with respect to the line-of-sight plotted as a function of height. These values are derived from the extrapolated magnetic fields at the loop footpoint marked by a black cross in Figs.~\ref{fig1}d \& \ref{fig1}e. The vertical dotted lines correspond to the section of the loop over which the propagation speeds are estimated. b) Actual propagation speeds along the loop after deprojection using the derived inclination angles. c) Temperature profiles along the loop as derived from the isothermal propagation of the waves in both channels. Dashed lines show the corresponding values calculated for adiabatic propagation. Measurement errors propagated from the observed phase speeds are also shown in b) and c). Note that the x-axis in b) and c) displays the actual distance along the loop, rather than the projected distance measured from the images.}
\label{fig4}
\end{figure}

The plasma $\beta$ (ratio of gas pressure to magnetic pressure) in the solar corona is usually very low ($\ll$1), which guides the slow magneto-acoustic waves to propagate at the local sound speed. However, the measured propagation speeds are normally along projections of the loop onto a 2-D image plane perpendicular to the line-of-sight. Thus, the observed speed, $v_{obs}$, is related to the sound speed, $c_{s}$, as
\begin{equation}
\label{eq1}
 v_{obs}=c_{s }~\mathrm{sin}{\,}\theta=\sqrt{\frac{\gamma \mathrm{RT}}{\mu}}~\mathrm{sin}{\,}\theta	,	
\end{equation} where $\gamma$ is the polytropic index, T is the plasma temperature, R(=8.314$\times 10^{7}$ erg{\,}K$^{-1}${\,}mol$^{-1}$) is the gas constant, $\mu$(=0.61) is the mean molecular weight \citep[mean mass per particle;][]{1993str..book.....M}, and $\theta$ is the angle of inclination of the loop with respect to the line-of-sight. The observed speed is essentially a function of the plasma temperature and the inclination angle of the loop. Note that, in principle, any variations in the polytropic index $\gamma$, and some non-linear effects could as well influence the observed propagation speed. However, the contribution of these effects is likely to minimal in the present case since the observed speeds change smoothly over distance and the wave amplitudes appear to be within linear regime. We use vector magnetograms of the photosphere, obtained by the Helioseismic and Magnetic Imager \citep[HMI;][]{2012SoPh..275..229S}, and employ nonlinear force-free field extrapolations \citep{2012ApJ...760...47G} to estimate inclination angles at different positions along the loop. Such extrapolations have proven to be robust in previous studies of this active region \citep{2013ApJ...779..168J,2016NPhys.12..179J}. 
The obtained inclination angles, with respect to the line-of-sight, are shown in Fig.~\ref{fig4}a as a function of height above the photosphere. The estimations were made at the loop footpoint marked by a cross in Figs.~\ref{fig1}d \& \ref{fig1}e. Considering the footpoint to be about 2 Mm above the photosphere, and taking the changing inclination angle into account, the height range between the vertical dotted lines in Fig.~\ref{fig4}a was identified to correspond to the loop segment over which the propagation speeds are measured. Using the obtained inclination angles, we deproject the measured phase speeds and estimate the corresponding local plasma temperatures following Equation~\ref{eq1}. The results are shown in Figs.~\ref{fig4}b \& \ref{fig4}c, where the respective errors propagated from the measured phase speeds are also displayed. Note that the x-axes in these plots show actual distances along the loop from the footpoint, rather than the projected distance used in Fig.~\ref{fig2}. Due to high thermal conduction in the solar corona \citep{2011ApJ...727L..32V}, and the short oscillation period (180 s), it may be appropriate to assume an isothermal propagation of the wave \citep{2004ApJ...616.1232K} under a linear approximation. However, the actual propagation could be isothermal or adiabatic, or somewhere in between, depending on the local conditions. As a result, we use $\gamma$=1 in these calculations, but also show the corresponding adiabatic solutions (for $\gamma$=5/3) as dashed lines in Fig.~\ref{fig4}c.

Clearly, the temperature in both channels increases with distance along the loop. Furthermore, there is an appreciable difference in the temperature between the two channels at all spatial locations along the loop. Since the loop structure visible in both channels is congruent, these results imply a multi-thermal (and consequently multi-stranded) structure of the loop. Importantly, the obtained temperature profiles highlight two distinct components of its multi-thermal cross section. We would like to note that the obtained temperatures near the loop footpoints are shifted from the peaks, but still within the temperature response functions of their respective AIA channels. However, the footpoints are still observable with reasonable (albeit significantly lower) intensities. Excess emission (i.e., above what would be expected from the contributions related to the loop footpoint temperatures) found in these locations may be a consequence of the inevitable optically-thin integration of background/foreground emission, whose contribution may be higher near the footpoints. Nevertheless, the fact that our derivation of the multi-thermal temperature profiles is independent of any such contamination (since the phase speeds are independent of the background loop intensity), highlights the diagnostic power of our technique.

\subsection{Spatial damping}
\begin{figure}
\centering
 \includegraphics[scale=1, clip=true]{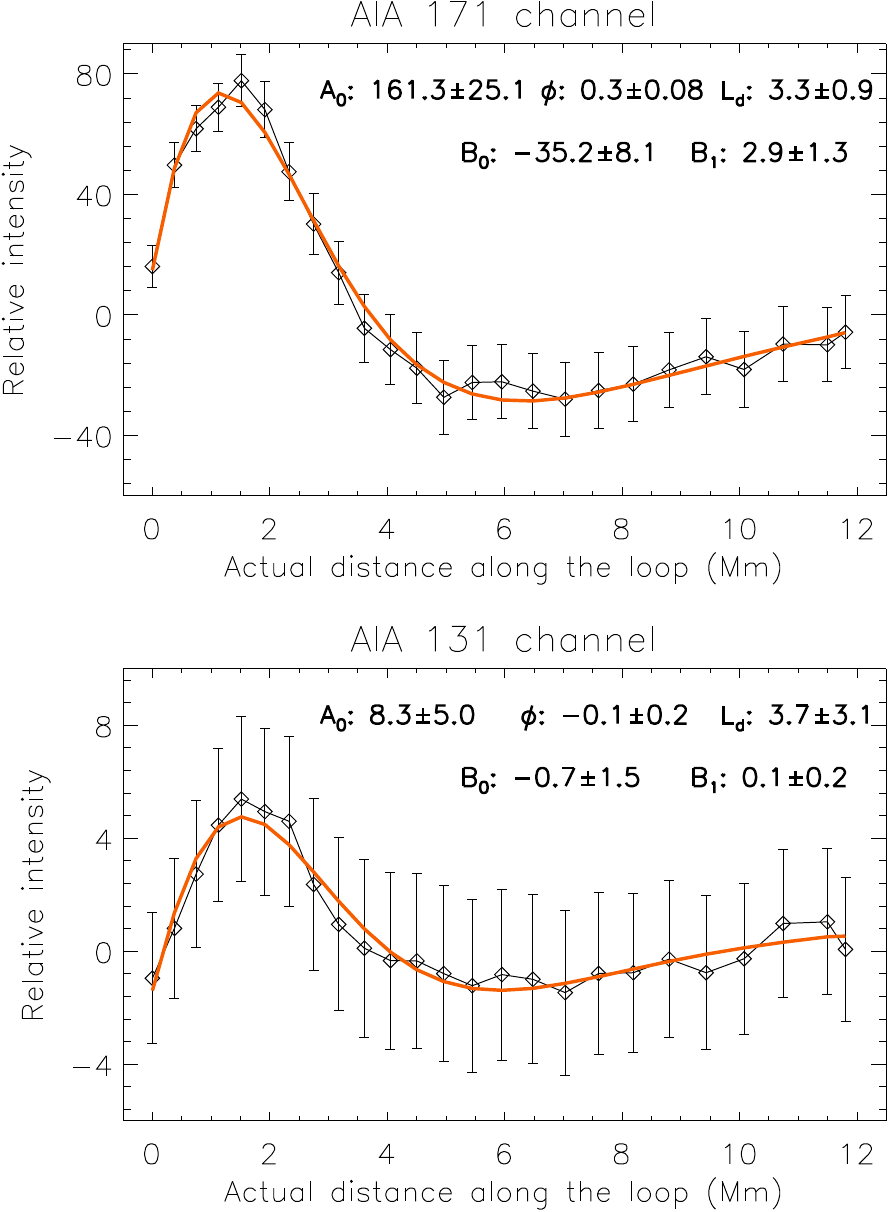}
 \caption{a) Spatial variation of intensities (background subtracted) from the 171{\,}\AA\ channel as a function of the (actual) distance along the loop, highlighting the damped propagation of the wave. The values correspond to the temporal location marked by the red arrows in Fig.~\ref{fig1}f and for the spatial extent bounded by the horizontal dashed lines in that figure. Errorbars represent errors in the values estimated from noise within the data. The overplotted red curve represents the best-fitting damped sinusoid with variable wavelengths following the function defined in Equation~\ref{eq2}. The obtained best-fit parameters are listed in the plot. b) Same as a) but for 131{\,}\AA\ channel.}
\label{fig5} 
 \end{figure}

Slow magnetoacoustic waves are known to damp as they propagate in the solar corona. As can be seen in Fig.~\ref{fig1}f, the waves can no longer be detected after a certain distance along the loop. To study this behavior and estimate the characteristic damping length, we analyze the spatial variation of intensity along the loop due to the wave. 
The intensities (background subtracted) from a particular instant in time as marked by red arrows in Fig.~\ref{fig1}f, occurring between the horizontal dashed lines is subsequently plotted in Fig.~\ref{fig5} for both channels. Note that the distances shown in this figure correspond to the actual (i.e., non-projected) values along the loop from the footpoint. The errors on the intensities have been estimated from noise contained within the data \citep{2012SoPh..275...41B,2012A&A...543A...9Y} from each of the respective channels. The red curves overplotted on the data represent the best fitting \citep{2009ASPC..411..251M} damped sinusoid, as defined by the function
\begin{equation}
\label{eq2}
I(x)=A_{0}{\,}e^{\left(\frac{-x}{L_{d}}\right)} \mathrm{sin}\left( \frac{2 \pi x}{\lambda (x)} + \phi \right) + B_{0}+B_{1} x.
\end{equation}Here, $x$ is the distance along the loop, $L_{d}$ is the damping length, $\lambda$ is the wavelength, $A_{0}$ and $\phi$ are the initial amplitude and phase, and $B_{0}$ and $B_{1}$ are appropriate constants. Since the phase speed of the observed wave is changing with distance as it propagates along the loop, its wavelength, $\lambda$, also varies as a function of distance. We estimate $\lambda (x)$ from the derived phase speeds and the oscillation period (180 s) for best-fitting the data. The obtained best-fit parameters are listed in the plot. The damping lengths are 3.3$\pm$0.9 Mm and 3.7$\pm$3.1 Mm for the 171{\,}\AA\ and 131{\,}\AA\ channels, respectively. The relatively short damping length in the hotter 171{\,}\AA\ channel is compatible with the theory of damping due to thermal conduction \citep{2012A&A...546A..50K}. However, this trend is not consistent at all temporal locations and the larger errors associated with the weaker 131{\,}\AA\ channel make it difficult to indisputably conclude. Nevertheless, the positive damping lengths (i.e., a reduction in amplitude) obtained in both channels are consistent with the presence of positive temperature gradients observed along the loop \citep{2004ApJ...616.1232K}. Using the temperature gradients, the response functions of the AIA 171{\,}\AA\ and 131{\,}\AA\ channels, and the assumption of hydrostatic equilibrium, we conclude that the observed damping is due primarily to wave dissipation and not an observational effect associated with temperature and density stratification \citep{2004ApJ...616.1232K}. Furthermore, the remarkable correspondence between the derived wavelength profiles, $\lambda (x)$, and the observed intensities (as visible from the quality of the fit), emphasizes the credibility of the derived phase speeds. Note that the increasing phase speed would also result in a reduction in the wave amplitude as a function of distance, which should also be considered while inferring any dissipation mechanisms from the observed amplitudes. In addition, it may not always be approrpriate to assume an exponential decay in wave amplitude since it is affected by variations in background physical conditions (e.g., local sound speeed) that are not necessarily exponential.

\section{Conclusions}
We present here the first ever unambiguous observations of multiple accelerating slow magnetoacoustic waves along a coronal loop. Evidence for small speed increases has been noted before \citep{2011A&A...528L...4K,2012SoPh..279..427K}, but was generally attributed to projection effects associated with a variable inclination of the structure rather than a true acceleration. However, in the present case, the change in speed is substantial even after accounting for the projection effects. The distinct speeds observed in the two channels indicate the multi-stranded and multi-thermal nature of the loop. The positive temperature gradients derived from the propagation speeds are consistent with the positive damping lengths obtained and are suggestive of more uniform heating along the loop \citep{1978ApJ...220..643R}. Moreover, these results are free from any possible contamination from overlapping structures in the line-of-sight \citep{2003A&A...406.1089D,2010A&A...515A...7T} and thus demonstrate the potential of MHD waves in revealing the basic physical properties of coronal loops. 

\acknowledgements 
The authors thank the anonymous referees for useful comments. S.K.P. wishes to thank S. Tomczyk for sharing his speed estimation software. D.B.J wishes to thank the UK Science and Technology Facilities Council (STFC) for the award of an Ernest Rutherford Fellowship in addition to a dedicated research grant. J.A.K. would like to thank the NASA Supporting Research and Technology Program for support. 


\end{document}